\documentclass{iau}
\usepackage{graphicx}
\usepackage{wrapfig}

\title[IAUS291.~~Progenitors and Environs of HBPs and magnetars] 
{On the environments and progenitors of SNRs associated with highly magnetized neutron stars} 
\author[Safi-Harb \& Kumar]
{S. Safi-Harb$^1$
 \and H. S. Kumar$^2$}

\affiliation{Department of Physics \& Astronomy, University of Manitoba, Winnipeg, Canada \\ $^1$email: {\tt samar@physics.umanitoba.ca, Canada Research Chair \\ 
$^2$ harsha@physics.umanitoba.ca} \\[\affilskip]}

\pubyear{2012}
\volume{291}  
\jname{\mbox{Neutron Stars and Pulsars: Challenges and Opportunities after 80 years}}
\editors{J. van Leeuwen, ed.} 
\begin{document}

\maketitle

\begin{abstract}
 The distinction between the high-magnetic field pulsars (HBPs, thought to be mainly rotation-powered) and magnetars (commonly believed to be powered by their super-strong magnetic fields) has been recently blurred with the discovery of magnetar-like activity from the HBP J1846$-$0258 in the SNR Kes~75. What determines the spin properties of a neutron star at birth and its manifestation as a magnetar-like or more classical pulsar is still not clear. Furthermore, although a few studies have suggested very massive progenitors for magnetars, there is currently no consensus on the progenitors of these objects. To address these questions, we examine their environments by studying or revisiting their securely associated SNRs. Our approach is by: 1)~ inferring the mass of their progenitor stars through X-ray spectroscopic studies of the thermally emitting supernova ejecta, and 2) investigating the physical properties of their hosting SNRs and ambient conditions. We here 
highlight our detailed  studies of two SNRs: G292.2--0.5 associated with the HBP J1119$-$6127 and Kes~73 associated with the AXP 1E~1841$-$045, and summarize the current view of the other (handful) HBP/magnetar-SNR associations. 
\end{abstract}

\firstsection 

\section{High Magnetic Field Pulsars and Magnetars: On their link, progenitors, and association with Supernova Remnants}
The past decade has witnessed a synergy of X-ray and radio observations revealing a diversity of young isolated neutron stars with magnetic fields spanning five orders of magnitude. These include the magnetars (the anomalous X-ray pulsars (AXPs) and the soft gamma-ray repeaters (SGRs)),
high-magnetic field radio pulsars (HBPs), Rotating Radio Transients, 
and the Central Compact Objects in Supernova Remnants (SNRs), in addition to the
more `classical' rotation-powered pulsars like the Crab.
X-ray observations of the HBP J1846--0258 in the SNR Kes 75  showed the first evidence of a magnetar-like behaviour from a Crab-like pulsar, suggesting that the class of HBPs is linked to magnetars (Kumar \& Safi-Harb 2008, Gavriil et al. 2008).
More recently, a magnetar was discovered with a relatively low magnetic field (Rea et al. 2010), a surprising result questioning the need for a high dipolar field ($B$$\geq$$B_{QED}$=4.4$\times$10$^{13}$~G) for magnetar-like activity. This also suggests that these objects are more common than commonly believed and that there is likely an evolutionary link between the different classes of neutron stars.

While observationally there is so far no consensus on magnetars' progenitors,  there is accumulating evidence for  them originating from very massive progenitors ($M$$>$30$M_{\odot}$), 
as inferred for SGR~1806--20 and CXOU J164710.2--455216 associated with very massive clusters (Figer et al. 2005, Muno et al. 2006),
 the presence of a stellar-wind blown bubble around AXP 1E 1048.1--5937 (Gaensler et al. 2005),
 and the Wolf-Rayet progenitor inferred for
 Kes~75 hosting HBP J1846--0258 (Morton et al. 2007).
However a lower progenitor mass of 17 $M_{\odot}$ has been suggested for SGR~1900+14 (Davies et al. 2009).

To further address the  link between HBPs and magnetars and their progenitors, we investigate their associated SNRs in X-rays.
X-ray spectroscopy provides a powerful tool to infer the mass of the progenitor. This is achieved by fitting the X-ray spectra of young, ejecta-dominated, SNRs and comparing the fitted metal abundances to nucleosynthesis model yields.
Furthermore, the parameters inferred from fitting the X-ray spectrum of the blast wave yield other intrinsic properties of the SNR (explosion energy, age, ambient density), thus shedding light on their environment and evolutionary stage.

Currently, there is only a handful of associations:
two SNRs securely associated with HBPs: G292.2--0.5/J1119--6127 and Kes~75/J1846--0258,
and four SNRs associated with magnetars: Kes~73/1E~1841--045,  CTB~109/1E~2259+586, G327.2--0.1/1E~1547.0--5408, and G337.0--0.1/SGR~1627--41.
Other proposed associations are: 
W41/magnetar Swift J1834.9--0846, G29.6+0.1 and CTB~37B with magnetars candidates AX~J1845--0258 and CXOU J171405.7--381031, respectively,  the candidate SNR G333.9+0.0/radio magnetar  
PSR J1622--4950, G42.8+0.6/SGR~1900+14, G353.6--0.7/magnetar candidate XMMU J173203.3--344518, and N49/SGR 0526--66 in the LMC\footnote{see  http://www.physics.mcgill.ca/$\sim$pulsar/magnetar/main.html  for the magnetars catalogue, and http://www.physics.umanitoba.ca/snr/SNRcat for the Galactic SNRs 
high-energy catalogue, associations with compact objects, and  references not listed here due to space limit.}. 
 We here briefly summarize our dedicated studies of  G292.2$-$0.5 and Kes~73
 (Kumar et al. 2012a, b), and conclude with a summary of the properties of their associated SNRs studied in X-rays.

\section{The SNRs G292.2--0.5 and Kes~73}

The SNR G292.2--0.5 is associated with the HBP J1119--6127 which has a rotation period $P$$\sim$408 ms, a characteristic age $\tau$$\sim$1.6~kyr (with an upper limit on its age of 1.9~kyr), and a dipole magnetic field $B$$\sim$4.1$\times$10$^{13}$~G. 
The combined \textit{Chandra} and \textit{XMM-Newton} study shows that the plasma is best described by a two-component thermal+non-thermal model.
The thermal component is fitted with a non-equilibrium ionization
model with a high temperature $kT$ ranging from 1.3~keV in the western side to 2.3 keV in the east, a column density increasing from 1.0$\times$10$^{22}$~cm$^{22}$ in the west to 
1.8$\times$10$^{22}$~cm$^{-2}$ in the east, and a low ionization timescale ranging from  5.7$\times$10$^9$~cm$^{-3}$~s in the SNR interior to 
3.6$\times$10$^{10}$~cm$^{-3}$~s in the western side.
An additional hard non-thermal component for the eastern and western sides of the pulsar can be partly attributed to leakage of relativistic particles from the pulsar or its associated nebula.
The spatial and spectral differences across the SNR are consistent with the presence of a dark cloud in the eastern part of the SNR absorbing the soft X-ray emission. 
The metal abundances inferred for the western side of the SNR are consistent with solar or sub-solar values, characterizing the emission from the supernova blast wave; while the interior regions indicate the presence of slightly enhanced abundances from Ne, Mg, Si, hinting for the first time at the presence of reverse-shocked ejecta.  We infer a high progenitor mass of $\sim$30$M_{\odot}$ suggesting a type Ib/c supernova,  an
SNR age of $\sim$4--7~kyr, and a low ambient density (Table~1). The discrepancy between the SNR and pulsar's age can be attributed to a variable braking index for the pulsar which 
recently showed unusual timing characteristics in the radio.

For the SNR Kes~73 associated with AXP 1E 1841$-$045 ($P$$\sim$11.8~s, $\tau$$\sim$4.7~kyr,  $B$$\sim$7$\times$10$^{14}$~G), our 
 \textit{Chandra} and \textit{XMM-Newton} spatially resolved spectroscopic study
 requires a two-component non-equilibrium ionization thermal model.
The soft-component has temperatures $\sim$0.3--0.5 keV and ionization timescales $\gtrsim$10$^{12}$~cm$^{-3}$~s 
with enhanced metal abundances;
 while the hard-component exhibits plasma
  temperatures $\sim$1.1--1.7~keV and low ionization timescales $\sim$(0.5--2.8)$\times$10$^{11}$~cm$^{-3}$~s with solar abundances.
  These results indicate that the soft-component arises from the reverse-shocked ejecta with most regions showing plasma that has reached ionization equilibrium, while the hard-component originates from the blast wave shocking an ambient medium with an average density of $n_0$$\sim$0.5~cm$^{-3}$. 
 We infer a progenitor mass of $\sim$(25--30) $M_{\odot}$, supporting the earlier prediction of a SN type IIL/b for the remnant, and an SNR age of $\leq$2.1~kyr (Table~1).

\section{Summary}
In summary most current studies point to highly magnetized neutron stars originating from very massive progenitors.
Table~1 highlights the SNR properties of the X-ray studied associations.
These results also suggest young SNRs (age $<$10~kyr) expanding in a relatively low-density medium for the Galactic SNRs. 
Further dedicated studies in X-rays and other wavelengths, as well as improved nucleosynthesis models, are needed to confirm these calculations
and increase the sample of neutron star--SNR associations.

\begin{table}[h]
\caption{SNR--HBP/magnetars associations with SNR properties inferred from X-ray studies.}
\begin{tabular}{l l l l l l l}
\hline
SNR/PSR & $D$~(kpc) & $n_0$ (cm$^{-3}$) & $v$~(km~s$^{-1}$) & Age~(kyr) & $E_0$ (10$^{51}$~ergs) & Ref. \\
\hline
G292.2--0.5/J1119--6127 & 8.4 & 0.02$f^{-1/2}$ & 1100 & 4.2--7.1 & 0.6$f^{-1/2}$ & [1] \\
Kes 75/J1846--0258 & 10.6 & 0.2--1.2 & 3,700 & 0.6--0.9 & -- & [2]\\
\hline
Kes 73/1E 1841--045 & 8.5 & (0.3--0.9)$f^{-1/2}$ & 1000 & 1.1--2.1 & 0.2$f^{-1/2}$ & [3] \\
CTB 109/1E 2259+586 & 3.0 & 0.16 & 720--1140 & 7.9--9.7 & 0.7--1.8 & [4] \\
CTB 37B/CXOU~J1714--3810 & 10.2 & 0.2--0.4 & -- & 0.4--3.1 & -- & [5]\\
N49/SGR 0526--66 (LMC) & 50 & 1.9$f^{-1/2}$ & 700 & 4.8 & 1.8$f^{-1/2}$ & [6]\\
 \hline
\end{tabular}
{\it Notes:}
$f$ is the volume filling factor.
The lower and upper age estimates for G292.2--0.5 and Kes~73 correspond to 
the free expansion and Sedov phases, respectively. 
For the other SNRs, the age was mainly estimated assuming the Sedov phase
(for CTB~37B using the ionization timescale). 
References: [1] Kumar et al. 2012a, [2] Temim et al. 2012, Su et al. 2009, [3] Kumar et al. 2012b,
[4] Sasaki et al. 2004, [5] Nakamura et al. 2009, [6] Park et al. 2012.  \\
\end{table}

\vspace{-0.3cm}
SSH acknowledges the support of the CRC program, NSERC, CFI, CITA, and CSA.
Thanks to Eric Gotthelf and Gilles Ferrand for comments.

\end{document}